\documentclass[psfig,useAMS,usenatbib]{mn2e}
\usepackage{aas_macros}
\usepackage{graphicx}
\usepackage{epstopdf}
\usepackage{psfig}
\usepackage{multirow}
\usepackage{url}
\usepackage{amssymb}

\title[Model Breaking]{Figures of Merit for Testing Standard Models:\\ Application to Dark Energy Experiments in Cosmology}
\author[Amara \&  Kitching ]{A. Amara\thanks{adam.amara@phys.ethz.ch}$^{1}$ \& T. D.  Kitching\thanks{tdk@roe.ac.uk}$^{2}$ \\
$^{1}$Department of Physics, ETH Zurich, Wolfgang-Pauli-Strasse 16,
  CH-8093 Zurich, Switzerland\\
$^{2}$SUPA, Institute for Astronomy, University of Edinburgh, Royal Observatory Edinburgh, Blackford Hill, EH9 3HJ\\
}

\newcommand{\be}{\begin{equation}}
\newcommand{\ee}{\end{equation}}
\newcommand{\ba}{\begin{eqnarray}}
\newcommand{\ea}{\end{eqnarray}}

\def\gs{\mathrel{\raise1.16pt\hbox{$>$}\kern-7.0pt %
\lower3.06pt\hbox{{$\scriptstyle \sim$}}}}         %
\def\ls{\mathrel{\raise1.16pt\hbox{$<$}\kern-7.0pt %
\lower3.06pt\hbox{{$\scriptstyle \sim$}}}}         %

\newcommand{\F}{\mbox{\boldmath $F$}}

\def\gs{\mathrel{\raise1.16pt\hbox{$>$}\kern-7.0pt %
\lower3.06pt\hbox{{$\scriptstyle \sim$}}}}         %
\def\ls{\mathrel{\raise1.16pt\hbox{$<$}\kern-7.0pt %
\lower3.06pt\hbox{{$\scriptstyle \sim$}}}}         %

\date{Accepted ---. Received ---; in original form ---.}

\pubyear{2009}

\begin{document}

\maketitle

\label{firstpage}

\begin{abstract}
Given a standard model to test, an experiment can be designed to: (i)
measure the standard model parameters; (ii) extend the standard model;
or (iii) look for evidence of deviations from the standard model. To
measure (or extend) the standard model, the Fisher matrix is widely
used in cosmology to predict expected parameter errors for future
surveys under Gaussian assumptions. In this article, we present a
framework that can be used to design experiments such that it maximises
the chance of finding a deviation from the standard model. Using a
simple illustrative example, discussed in the appendix, we show that
the optimal experimental configuration can depend dramatically on the
optimisation approach chosen. We also show some simple cosmology
calculations, where we study Baryonic Acoustic Oscillation and
Supernove surveys. In doing so, we also show how external data, such as
the positions of the CMB peaks measured by WMAP, and theory priors
can be included in the analysis. In the cosmological cases that we
have studied (DETF Stage III), we find that the three optimisation
approaches yield similar results, which is reassuring and indicates
that the choice of optimal experiment is fairly robust at this level. However, this may not be the case as we move to more ambitious future
surveys.  
\end{abstract}


\section{Introduction}

In cosmology, the $\Lambda$CDM concordance model has become our
standard model of the Universe. This model satisfies
current data and depends on three critical sectors: (i) Dark Energy;
(ii) Dark Matter; and (iii) Initial Conditions. These sectors are linked through
our theory of gravity - general relativity. Although this model is well
defined, the addition of each component has typically been done to
explain the available data rather than arising from some fundamental
theory of the cosmos. Hence, cosmology is currently in a data-driven era,
with little known about the fundamental nature of dark
matter and dark energy. 
As a result, a significant effort is underway
in this very active field to build experiments to measure and extend our standard model. These include KIDS, Pan-STARRS\footnote{\url{http://pan-starrs.ifa.hawaii.edu}},  
DES\footnote{\url{http://www.darkenergysurvey.org}}, 
LSST\footnote{\url{http://www.lsst.org}},
JDEM\footnote{\url{http://jdem.gsfc.nasa.gov}} and Euclid\footnote{\url{http://www.euclid-imaging.net}}$^,$\footnote{\url{http://sci.esa.int/euclid}}. In planning such future
observations, the approach to date has been to optimise the
experimental and methodological designs to minimise the errors on
extended parameters. In particular, the dark energy equation of state (the ratio of pressure to density of dark energy
$w(z)$) garners the most attentions and is typically parameterised in terms
of a second order Taylor expansion in the scale factor or redshift
z (e.g. $w(z)=w_0+w_a z/(1+z)$). Experiments are then designed to measure these equation of
state parameters to the highest possible precision. The
dark energy Figure of Merit \citep[FoM; ][]{2006astro.ph..9591A}, which is proportional to the
area of the error ellipse in the $w_0$-$w_a$ plane is widely used to
gauge performance. Other possible metrics have also been suggested,
such as the addition of parameters to test for deviations from
Einstein gravity or the division of $w(z)$ into a large number of
redshift slices that can then be used to construct principal
components through a matrix inversion \citep{2009arXiv0901.0721A,
  2003PhRvL..90c1301H}. However, these two suffer from 
their own problems. For instance, the additional modified gravity
parameters may not be strongly motivated and the eigenfunction
decomposition of $w(z)$ can suffer from instabilities
\citep{2009MNRAS.398.2134K}. 

In this article, we present an alternative methodology to be applied to experimental
design when faced with a standard model and no guidance from
theory. We show that an experiment can be designed such that the
probability of breaking the standard model (finding evidence against
the model) can be maximised.   

This article is organised as follows. In Section \ref{Approaches to
  Experiment Designing}, we review the alternative approaches to
experimental design. We then, in Section \ref{application}, compare each approach using a simple explanatory model, as well as
a cosmological example that studies the performance of the `current' and
Stage III experiments discussed in \cite{2006astro.ph..9591A}. We summarise our
conclusions in Section \ref{Conclusion}.  

\section{Approaches to Experiment Designing}
\label{Approaches to Experiment Designing}

When planning an experiment with a standard model (a set of
parameters) in mind, we can think of three possible approaches that we can take. The first is to stay within the standard model and to design
an experiment that will measure the parameters of this model to the
highest possible precision. The next is to extend the standard model
(add extra parameters), and ideally this extension would be driven by a
compelling theoretical framework with clear testable
predictions. Finally, in the absence of any compelling theory, one can
take a more exploratory approach, where the driving aim is to design an
experiment with the greatest chance of breaking the standard
model. Ideally, this approach would depend only on well-founded
knowledge, such as today's data, the expected error bars of future data and
the standard model that is being tested.  

\subsection{Measuring the Standard Model}
\label{sec:stdmod}
Within a well-specified model, the Fisher matrix formalism \citep{1997ApJ...480...22T} is a
well-defined framework for estimating the errors 
that a given experiment will have on the measurement of the parameters of
the model. For an experiment where the parameters have an effect on
the  mean, the Fisher matrix is defined as  
\be
F_{ij} = \sum{\frac{1}{\Delta C^2} \frac{\partial C}{\partial
    \Theta_i} \frac{\partial C}{\partial \Theta_j}}, 
\ee
where $C$ is some observable signal, $\Delta C$ is the expected error
for an experiment and $\Theta$ is a vector containing the
parameters. A cosmology model may include $\Theta = \{\sigma_8,
~\Omega_m, ~\Omega_b, ~\Omega_\Lambda, ~n_s, ~h, {\rm etc}$\}, where, for
instance, the dark energy equation of state is assumed to be a
cosmological constant ($w(z)\equiv -1$).  The errors on each of these
parameters are then given by the diagonal elements of the parameter
covariance matrix ($Cov$), which is given by $Cov=F^{-1}$. 

\subsection{Extending the Standard Model}
\label{sec:ext_stdmod}
When seeking out new physics, we look for ways of going beyond the
standard model. Ideally this would be done through the guidance of
theory. There are many examples of cases where theories have been put
to the test by experiments based on verifiable predictions. One such
example is neutrino mass. In the standard model of particle physics,
neutrinos have zero mass, but the assumption of zero mass is an ad hoc choice. A natural and physically motivated extension of this
model was to add mass to neutrinos (through the lepton mixing matrix
addendum). Neutrino mass has now been experimentally confirmed by a
number of particle physics experiments 
\citep{2004PhRvL..92r1301A,2003PhRvL..90b1802E,2006PhRvD..74g2003A}, and cosmological experiments should be able to
constrain this mass to high accuracy \citep[e.g.][]{2010arXiv1001.0061R,
  2009arXiv0911.5291T, 2008PhRvD..77j3008K}. 

Extra parameters, $\Psi$, can be added to the parameters of the
standard model, $\Theta$. In this case, the Fisher matrix formalism 
can once again be used to
estimate the errors on all the parameter sets. Here, it becomes useful
to decompose the matrix as  
\begin{equation}
F  = \left(\begin{array}{cc} F^{\Theta\Theta} & F^{\Theta\Psi}
  \\F^{\Psi\Theta} & F^{\Psi\Psi}\end{array}\right), 
\end{equation}
where the matrix $ F^{\Theta\Theta}$ contains the Fisher matrix
elements for the parameters of the standard model, $F^{\Psi\Psi}$
contains the elements for the new model parameters and
$F^{\Theta\Psi}$ contains the cross terms.   

This approach has been widely adopted by the cosmological community in
dark energy studies. In this case, the extra parameters are typically
added in the form of equation of state parameters (the ratio of
pressure to density) of dark energy ($w$). However, this is a specific
way of thinking about dark energy (as a dynamical fluid). Therefore, models
that do not treat dark energy as a fluid have to work in terms of an
`effective' equation of state. A further complexity arises because the
observed low redshift acceleration that motivates dark energy could result from other physics, such as the breakdown of Einstein gravity on
cosmic scales. A move away from Einstein gravity may not be well
represented by the addition of equation of state parameters and may require
the addition of new parameters that specifically allow for such
deviations. As a result, these extra dark energy parameters do not have
a firm theoretical basis but are, in fact, an arbitrary expansion of the
equation of state \citep{2009MNRAS.398.2134K}.  

\subsection{Breaking the Standard Model}

Here, we introduce a new approach to experimental planning, where 
we explicitly design an experiment to maximise the probability of finding a
deviation from the standard model. This deviation is allowed to come
from any part of the theory and should not depend on any particular
theoretical extension of the standard model. The robustness of such an
approach can be achieved by relying on minimal inputs, namely: (i)
current data; (ii) expected error bars of future measurements; and
(iii) the standard model that we want to test.  

We begin by defining some basic parameters. Let $X$ be a data vector
containing today's measurements (for instance a correlation
function). These data points have associated errors, $\sigma_X^2$,
which means that the measured data points are randomly scattered about T, the data vector that would be measured with no measurement error or systematic, 
i.e. the underlying values of the observable as measured with
the perfect experiment\footnote{As an example, if X is calculated from the mean of $n$ independent data points and the errors are given by the variance ($\sigma^2(\bar{X}) = \sigma^2(X)/n^2$), then T would be the measure given as n goes to infinity in the absence of systematics. We note that, in this case, cosmic variance would come from the fact that due to a finite Universe the number of independent data points will be limited to a finite number.}. 
The expected error bars of a future experiment
are $\sigma_Y^2$, which would produce a data vector $Y$. Given today's data, we can
calculate the probability of the future data,
$P(Y|X)$, by marginalising over T,   
\begin{equation}
\label{eq:pyx}
P(Y|X) = \int P(Y|T) P(T|X)~ dT,
\end{equation}
where $P(T|X)$ is the probability of T given today's data and
$P(Y|T)$ is the probability of the future data given T. The
integral is performed over all possible T since we do not know
what T is a priori. 

For each realisation of the future
data, there will be an associated best-fit that can be achieved with
the standard model. We focus here on the $\chi_{\rm min}^2$. With the probability
distribution of future data given current data ($P(Y|X)$), which, for simplicity, we will sometimes also denote using $P(Y)$, we can calculate the expectation
value of the minimum $\chi^2$ by integrating over all possible future
data vectors:   
\begin{equation}
\label{eq:chi}
\langle \chi^2_{\rm min} \rangle = \int \chi^2_{\rm min}(Y) P(Y)~dY. 
\end{equation}
A high $\chi^2_{\rm min}$ means that the standard model is not able to
give a good fit to future data. Hence, an experiment designer who
wants to maximise his or her chances of breaking the standard model should
focus on an experiment configuration that maximises the expectation value of the
minimum $\chi^2$; ${\rm max}[\langle\chi^2_{\rm min}\rangle]$.
Strictly, we should use a quantity that is robust to the number of
data points (for instance the reduced $\chi^2$). We avoid such
problems in what follows by only making comparisons between
experiments with equal numbers of data points. The $\chi^2$ and reduced
$\chi^2$ are, therefore, simply scaled versions of each other. In this work, we have focused on the expectation value of the minimum $\chi^2$ of the future data, with the understanding that a $\chi^2$ corresponding to a reduced $\chi^2$ significantly larger than one will require additional parameters beyond those available in the standard model. However, it may be interesting to also consider the higher order statistics of the minimum $\chi^2$ distribution. Along similar lines of thought, our FoM could also be recast in terms of the probability that a future experiment will give a $\chi^2_{\rm min}$ greater than some threshold value. For the work presented here, we use the simplest expression (given in equation \ref{eq:chi}), but we are continuing to investigate further possible expressions of this model breaking FoM.

Here, we use the maximum likelihood fit to the data (minimum
$\chi^2$). We have used this frequentist measure, as opposed to a
Bayesian evidence criteria, because there are no 
objective Bayesian measures in the case of assessing the quality of a theoretical fit for a single model, given that a single model Bayesian evidence must conclude (through a normalisation of probabilities) that
there is $100\%$ evidence for that model (see Taylor \& Kitching, 2010
for further discussion). In general, this $\chi^2_{\rm min}(Y)$ measure
could be replaced with any `goodness of fit' criteria $G(Y)$, where equation \ref{eq:chi}
optimises fit.

\section{Application}
\label{application}
\subsection{Illustrative Example}

In Appendix A, we explore the impact of the choice of optimisation
metric on a simple illustrative example. We set up a system of three
data points and `a standard model' that is a straight line with one
degree of freedom - the slope of the line. What this shows is that the
optimal configuration of a future experiment can vary drastically and
can lead to exactly opposite optimisations in some cases depending on
whether model breaking or standard model extension is used.   

The simple model that we set up has a `pivot point,' where the model
makes an exact prediction, $C(x=8)\equiv 10$. To measure the standard
model parameter (the slope), assuming that this model is correct, it is clear that there is no
sensitivity at this point. Therefore, an optimisation will minimise future error bars
away from the pivot point. However, in the model breaking mode, it is
optimal to place the smallest future error bars at the pivot point,
since it is here that even the slightest deviation from the standard
model prediction would yield proof that the standard model is
broken. Of course the model breaking paradigm here is a high-risk,
high-gain approach. If T happens to have the same value as that of the pivot point, then this approach would yield no extra
information. When extending the standard model, the optimal
configuration is entirely dependent on the exact form of the
extension. For instance, a clear difference is seen between a standard
model that is extended by adding a constant parameter and one that is
expanded with a parabolic term about the pivot point, thereby preserving the pivot point.  

\subsection{Cosmological Application}

We now apply our approach to investigate the planning of cosmology
surveys. In this work, we focus on some simple examples that show how
this can be done, with a more complete investigation of future surveys
to follow in later work. In this example, we focus on supernovea (SNe) \citep{1998astro.ph..5117T} and Baryon Accoustic Oscillation (BAO) \citep[e.g. see][for discussion]{2008arXiv0810.0003R}. In
addition, we will show: (i) how external data, in this case the CMB peak
separation, can be added; (ii) how priors coming from theory can be
included; and (iii) a simple treatment for systematics errors.     

\subsubsection{Survey Configurations}

Due to the computational limits of performing the integral shown in equation \ref{eq:chi}, the dimensionality of which scales with the number of data points, we have decided to bin the low redshift data (i.e. SNe and BAO) into four redshift bins (i. $0.1 < z< 0.4$; ii. $0.4 < z<0.7$; iii. $0.7 <1.0$; and iv. $1.0 <z< 1.3$). By fixing the number of redshift bins, and therefore the number of degrees of freedom since the standard cosmology model is the same for all cases, we are able to compare the $\chi^2_{\rm min}$ values directly. This simplifies the comparison between different survey configurations. For current BAO data, we use the galaxy number counts presented in \cite {2010MNRAS.401.2148P}. This work presented a BAO analysis of the Sloan Digital Sky Survey Data Release 7 sample (DR7).  This is composed of roughly 900,000 galaxies over 9100 deg$^2$ in the redshift range z = [0.0, 0.5]. We re-binned this data into our four redshift bins which leads to the distribution shown in Table \ref{tbl:surveys1}. For current SNe data, we use the Union data presented in \cite{2008ApJ...686..749K}. This is a compilation of SNe data coming from a number of measurements, including the Supernova Legacy Survey, the ESSENCE Survey and supernovae measurements from the Hubble Space Telescope (HST). Once again, as with the BAO data, we have re-binned this data to match the four bins that we use in this paper (see Table \ref{tbl:surveys2}). As we will discuss in Section \ref{sec:extdata}, we have also included constraints coming from current measurements of the CMB peak separation presented in \cite{2009ApJS..180..330K}, which uses the WMAP data.

For future surveys, we have decided to focus on a configuration that illustrates the technique presented here, rather than to make concrete recommendations about specific mission concepts. The reason for this is that the calculations that we present here include a number of simplifications, such as using only four redshfit bins. These, we feel, allow us to calculate trends and make some statements about the relative merits of broad concept ideas. However, to draw detailed conclusions on specific mission configurations would take further detailed work that we will address in follow up publications on this topic.  For the future surveys that we use to illustrate our method we have relied on the Stage III surveys described in \cite{2006astro.ph..9591A}, although many of the projects may have
evolved since this document was released. Once again, we re-bin the Stage III data into our four redshift bins (see Tables \ref{tbl:surveys1} and \ref{tbl:surveys2}).  

For the BAO surveys, we simplify the analysis by
only using the tangential modes, which is pessimistic, and assume no
systematics, which is optimistic. Due to these reasons, the results
below are illustrative, and we do not claim that the optimistic and
pessimistic approaches cancel out each other. We calculate the errors on
BAO scale using the fitting function given in \cite{2006MNRAS.365..255B}, which has
been implemented in {\tt iCosmo} \citep{2008arXiv0810.1285R,2009arXiv0901.3143K}. For the Supernova error
calculations, we have used the Fisher matrix approach outlined in \cite{1997ApJ...480...22T} and \cite{2001PhRvD..64l3527H} and have assumed a
systematic contributions outlined in \cite{2004MNRAS.347..909K} and \cite{2006PhRvD..74d3513I}. However, we will also show results without systematics in order
to gauge their impact.

\begin{table*}
  \centering 
\begin{tabular}{|l|c|c|c|c|c|c}
\hline
& Area & \multicolumn{4}{|c|}{Number Density of Galaxies (${\rm n_g}$) [num/amin$^2$]} & \\
& & 0.1$<$z$<$0.4&0.4$<$z$<$0.7&0.7$<$z$<$1.0&1.0$<$z$<$1.3 \\
\hline
 Current  & 10000 & 0.013 & 0.00056&0.0&0.0 \\
 Stage III WiggleZ  & 1000 & 0.0&0.022&0.089&0.0\\
 Stage III BOSS &  10000 &  0.014&0.019&0.0&0.0\\
 Stage III WFMOS & 2000 & 0.0&0.056&0.22&0.18\\
\hline
\end{tabular}
  \caption{Parameters of the BAO surveys considered in this study. The current survey is chosen to be close to the BAO survey parameters for the SDSS DRL7 \citep{2010MNRAS.401.2148P}. The future surveys have been chosen from the Stage III surveys of the Dark Energy Task Force report \citep{2006astro.ph..9591A}.}
	\label{tbl:surveys1}
\end{table*}

\begin{table*}
  \centering 
\begin{tabular}{|l|c|c|c|c|c|}
\hline
&  \multicolumn{4}{|c|}{Number of Supernovae (${\rm n_s}$)} & \\
& 0.1$<$z$<$0.4&0.4$<$z$<$0.7&0.7$<$z$<$1.0&1.0$<$z$<$1.3 \\
\hline
Current SNe  & 51&107&131&18 \\
Stage III SNe & 965&1940&860&57 \\
\hline
\end{tabular}
  \caption{Parameters of the Supernovae surveys considered in this study. The current survey is chosen to be close to the Union supernovae sample  \citep{2008ApJ...686..749K}. The future surveys have been chosen from the Stage III surveys of the Dark Energy Task Force report \citep{2006astro.ph..9591A}.}
	\label{tbl:surveys2}
\end{table*}

\begin{figure}
\centering
\resizebox{\columnwidth}{!}{\includegraphics{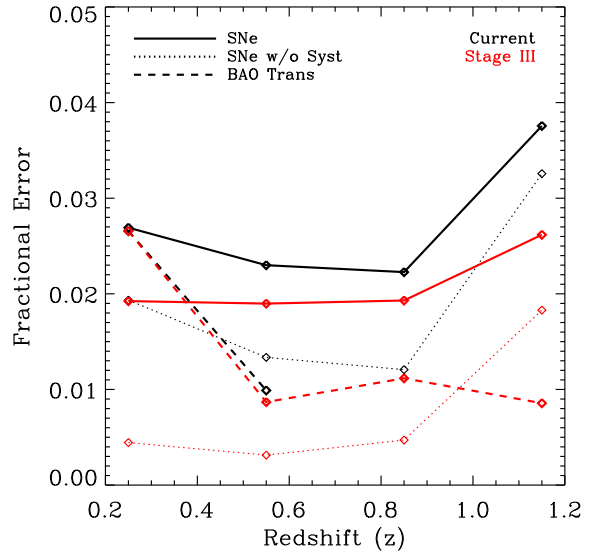}}
\caption{Fractional errors on the observed quantities for `current' (black) and stage III (red) experiments. For the BAO measurements, these are the errors on the transverse BAO scale from Blake et al. For the SNe surveys, the observable is the flux loss of the SNe.}
\label{fig:probeerrors} 
\end{figure}

\subsubsection{Including External Data}
\label{sec:extdata}
In this study, we focus on the potential of future BAO and SNe surveys. It is, however, straightforward to include other data sets. To do this, we must decide whether to only include current measurements (for instance, in the case of the CMB to include WMAP data) or try and anticipate the joint impact of future measurement of that probe (for instance, to include predictions for Planck\footnote{http://www.rssd.esa.int/SA/PLANCK/docs/Bluebook-ESA-SCI(2005)1\_V2.pdf}). If the latter is desired, then the prescription for doing so follows the same logic as that used for the BAO and SNe calculations and would increase the data vectors ($F$ and  $X$) in equation \ref{eq:chi}. While conceptually simple, adding external data in this way can quickly lead to computational challenges, since the dimensionality of the integral scales the number of data points. The computation time for convergent results can diverge quickly, even using a simple Monte-Carlo integration scheme. 
To solve potential problems, we would either need to develop a sophisticated Monte-Carlo integration scheme with, for instance, importance sampling that is tailor made for this problem or try to reduce the number of data points by focusing on specific features of the external data that we wish to consider.
For instance, in the case of the CMB we can consider adding the peak position and height information rather than implementing the full correlation data ($C(\ell)$).

If we only add existing external data, then the calculation is greatly simplified, since the dimensionality of the integral in equation  \ref{eq:chi} remains the same. Instead, the external data is simply used when calculating the minimum $\chi^2$. In the work presented here, we have included the measured spacing of the acoustic oscillation peaks of the CMB, $\ell_A$,  which depends on the ratio of angular diameter distance to the sound horizon at photon decoupling epoch ($z_*$),
\begin{equation}
\ell_A = (1+z_*)\frac{\pi D_A(z_*)}{r_s(z_*)},
\end{equation}
where $D_A$ is the angular diameter distance and $r_s$ is the sound horizon. This peak spacing has been measured to be $\ell_A = 302.1 \pm 0.86$ for WMAP  \citep{2009ApJS..180..330K}, which gives an expression for $z_*$ in equation 66. For the sound horizon calculation, we follow the calculations presented in Appendix A of \cite{2007MNRAS.377..185P}.

\subsubsection{Theory Priors and Calculating Probabilities}

We now turn our attention to priors coming from our theory and how these can bound our results. For example, if we impose no knowledge at all about what we expect, then the PDFs for each of the data points in equation \ref{eq:pyx} are independent. A simple consequence of this is that the probability distribution for future data in bins where no current data exists ($P(F|X)$) will be flat between $-\infty$ and $\infty$. Inputing this PDF into equation \ref{eq:chi} would lead to a $\langle \chi^2_{\rm min}\rangle$ of infinity, which is not fully useful when comparing expected performances. One can view this result in two ways. The first is that a data purist (i.e. someone who wishes not to add any bounds from theory) would conclude that the best surveys are those that explore new regions where no measurements have yet been made. The alternative approach is to introduce some expectation from our knowledge of basic cosmological theory. Theory priors modify the PDFs of future data by imposing relationships between different data points. A simple addition is to impose a link between the angular diameter distance and the luminosity distance.

For the configurations shown in Table \ref{tbl:surveys1}, we immediately see that if we take no guidance from theory then we will be driven towards  WiggleZ and WFMOS (see Table \ref{tbl:surveys1}), since these two surveys will provide BAO measurements at redshifts that are currently not explored by current BAO experiments and, hence, have an expectation value of minimum $\chi^2$  of infinity. Once again, a data purist may argue that these surveys should, therefore, be our top priority. In contrast, another simple approach is to rely on the widely accepted relationship between angular diameter distance ($D_A$) and
luminosity distance ($D_L$) given by 
\begin{equation}
D_L = (1+z)^2 D_A.
\end{equation}
By explicitly adding this very weak prior from the theory, the probability of future data is modified (equation \ref{eq:pyx}) to 
\begin{equation}
P(Y|X) = \int P(Y_B|D_L)P(Y_S|D_L) P(D_L|X_B)P(D_L|X_S)~ dD_L,
\end{equation}
where $Y_B$ and $Y_S$ are the data vectors for future surveys for BAO and SNe (respectively) and $X_B$ and $X_S$ are the data vectors for today's surveys.  This PDF, therefore, includes a relationship between the SNe measurements and the BAO measurements at any given redshift. For what we present later, this relationship between distances is the only information that we impose from theory. However, a natural question is what would happen if the future data were to extend to redshifts that are not covered by either the BAO or the SNe data? A detailed exploration of this will be presented in follow-up work. Nonetheless, here we give a brief discussion of the basic principles. Once again, priors from theory can be used to impose relationships between different data points, which in turn modify the PDF of the future data. In particular, the question raised here would look for relationships between data points at different redshifts. This can be done by introducing an integral relationship between distance (co-moving - $D_c$) and the Hubble function, $H(z)$,
\begin{equation}
D_c = c \int_0^{z^\prime} \frac{dz^\prime}{H(z^\prime)},
\end{equation}
where c is the speed of light. Without resorting to the Friedmann equation, which links H(z) to density parameters of the matter-energy components of the Universe, we can place simple constraints on the functional form of H(z) that can be used to compute the probability of future data. For instance, an assumption that H(z) is a positive definite function over cosmic time would bound the comoving distance at a redshift of $z_i$ to be between the comoving distances at  $z_{i-1}$ and $z_{i+1}$, i.e. that of the redshifts on either side. Here the inclusion of the CMB, with $z \sim 1100$, becomes very useful. The advantage of this approach is that all knowledge from theory, including simple relationships, such as that between $D_L$ and $D_A$, must be included explicitly. This then allows us to decide explicitly what assumptions should be included.

\subsubsection{Computation of $\langle \chi^2_{\rm min} \rangle$}

For each realisation of the the future data (Y) we calculate the weighted average data, which is given by
\begin{equation}
X_c = \frac{\sigma_X^2  Y  + \sigma^2_Y X}{\sigma^2_X + \sigma^2_Y},
\end{equation}
where $X_c$ is the value of the combined data, Y and X are the future and current data values, and $\sigma$ are the associated errors. The errors on the combined data are
\begin{equation}
\sigma^2_c = \frac{\sigma_X^2 \sigma^2_Y}{\sigma^2_X + \sigma^2_Y}.
\end{equation}
The data vector $X_c$ can also contain external data for which there will not be corresponding future measurements. In this case, the data vector enteries that correspond to the external data have $X_c = X$ and $\sigma_c = \sigma_x$.
With this combined data vector, we then calculate $\chi^2$,
\begin{equation}
\chi^2 = \sum \frac{(X_c - M)^2}{\sigma^2_c},
\end{equation}
where the sum is over the entries of the data vector. In our case, this corresponds to a total of nine data points (BAO scale at four redshifts, SNe at four redshifts and the CMB peak spacing). For a given choice of cosmology parameters, M is the value given by the model. For each integration step, we use a minimiser to find the parameters that lead to the smallest $\chi^2$ value.

The Stage III surveys will look for deviations from the standard
$\Lambda$CDM concordance model. We consider the standard cosmological model as one with Gaussian initial conditions\footnote{See \cite{2004MNRAS.351..375A, 2010CQGra..27l4011D, 2010MNRAS.402..191P} for examples of how non-Gaussian initial conditions impact observables at low redshifts} following inflation, with scale-free perturbations ($n_s$ =1), where spatial curvature is allowed and dark energy is understood to come from the cosmological constant $\Lambda$ (i.e. $w = -1$).  Since we
only consider the distance-redshift measurements, we are sensitive to the following parameters of the model\footnote{We note that there is a weak dependence on $\Omega_b$ through $z_*$, but we have neglected this here since it has little impact on the results and only complicates the calculation.}: \{$\Omega_m$,
$\Omega_\Lambda$, $h$\}. The model breaking approach does not rely a adding further parameters beyond these well-understood ones and will test how likely it is that future experiments, based on today's data, would find any deviation from $\Lambda$CDM, including, for example, evidence for $w \neq -1$.

We perform the integral in equation \ref{eq:chi} over all possible realisations of the
future data, which corresponds to an eight dimensional integral (four
future BAO and four future SNe). For practical reasons to do with computational feasiblity, we use the simple Monte-Carlo integration technique outlined in section 7.7 of Numerical Methods \citep{1403886}. Here, a multidimensional integral (in our case, equation \ref{eq:chi}) can be expressed as

\begin{equation}
\int f dV \approx V\langle f\rangle \pm V \sqrt{\frac{\langle f^2 \rangle - \langle f\rangle^2}{N}},
\end{equation}
where the expectation values, denoted by the angular brackets, can be calculated by randomly sampling the function $f$ at positions $x_i$ with
\begin{equation}
\langle f \rangle \equiv \frac{1}{N} \sum_{i=0}^{N-1} f(x_i).
\end{equation}
The volume of the parameter space is denoted as V. This is set by the bounds of the integral, which we have choosen in such a way as to ensure that the integrand is vanishingly small at this limit.

\subsubsection{Results}

Performing survey optimisations for future experiments typically involves a trade-off between different configurations that compete for resources. A classic example is a trade-off between the depth and area of a survey for a fixed exposure time \citep[see ][for an example of this for weak lensing surveys]{2007MNRAS.381.1018A}. Another, more difficult and often controversial trade-off study, is to trade-off resources between different proposed probes. For instance, if due to limited resources it is not possible to support both SNe and BAO missions envisioned for stage III. A natural question might be - should we invest in one over the other? Or should scaled-down versions of each mission be pursued? This is a complex issue for a number of reasons, but the model breaking figure of merit, along with other FoMs, can help guide such decisions by quantifying the likelihood of finding a deviation from `the standard cosmological model'. For this reason, our first illustrative example focuses on a possible trade-off study between SNe and BAO stage III surveys. We note again that a thorough  treatment of such a trade-off is complicated. For instance, quantifying the impact of limited resources is significantly more complicated than that of limited observation times. We made a number of simplifying assumption, so the results stated here are only to illustrate the method rather than to offer concrete recommendations about one experiment over another. In this spirit, we will show results for the full Stage III surveys, as well as for the scaled down versions. We do not attempt to make a link between the scaled-down versions for a fixed set of resources, since this is well beyond the scope of this work. For scaling down the surveys, we have decided to fix the distributions in redshifts (i.e. the PDF of the number of SNe and galaxies as a function of redshift is fixed), and we vary an overall scaling. For BAO this corresponds to a change in survey area, and for SNe this corresponds to a reduction in the total number of SNe.

\begin{figure}
\centering
\resizebox{\columnwidth}{!}{\includegraphics{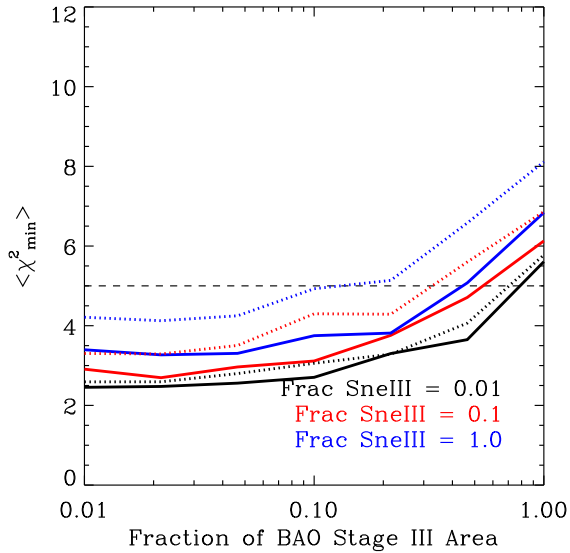}}
\caption{Expectation value of the minimum $\chi^2$ as a function of
  the areas of the stage III BAO surveys. The fraction corresponds to
  the fraction of the full survey areas (shown in Table
  \ref{tbl:surveys1}) 
  used. These are shown for three configurations of stage III SNe
  surveys, where only a fraction of the SNe in Table \ref{tbl:surveys1}
  are used. The solid curves include SNe systematics while the dotted
  curves do not. The dashed line shows the $\chi^2$ that would correspond to a reduced $\chi^2$ of 1.} 
\label{fig:bao_area} 
\end{figure}

\begin{figure}
\centering
\resizebox{\columnwidth}{!}{\includegraphics{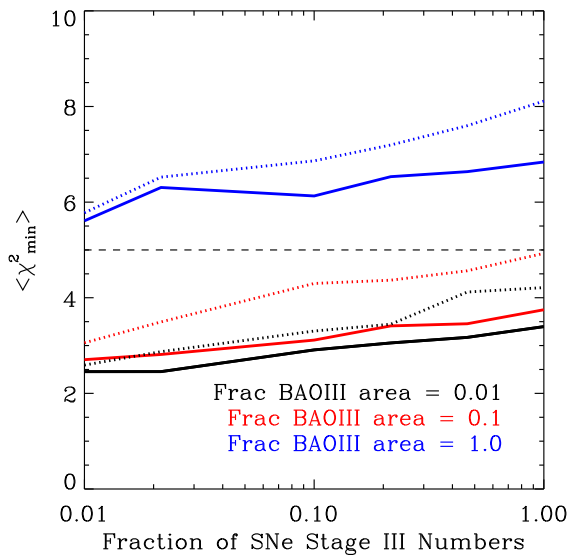}}
\caption{Expectation value of the minimum $\chi^2$ as a function of
  the number if SNe of the stage III surveys. The fraction corresponds
  to the fraction of the full survey number (shown in Table
  \ref{tbl:surveys1}) 
  used, where the PDF is fixed and only a global fraction is
  applied. These are shown for three configurations of stage III BAO
  survey area where only a fraction of the areas in Table
  \ref{tbl:surveys1} are 
  used. The solid curves include SNe systematics while the dotted
  curves do not. The dashed line shows the $\chi^2$ that would correspond to a reduced $\chi^2$ of 1.} 
\label{fig:sne_num} 
\end{figure}

In Figure \ref{fig:bao_area}, we show the expectation value of the
minimum $\chi^2$ when we consider only some fraction of the area of
the Stage III surveys shown in Table \ref{tbl:surveys1}. For instance, for a
fraction of $0.5$ we divide the areas of all the BAO missions by a
factor of $2$. The results are shown for different realisations of
Stage III SNe surveys, where once again the fraction refers to the
fraction of the total SNe numbers shown in Table \ref{tbl:surveys2}. 
We see that for a range of SNe stage III configurations increasing the area of the
BAO survey from 1\% to 10\% of what is expected in stage III has no effect on the expectation value of the minimum $\chi^2$. Beyond this, however, we see a large increase in $\langle \chi^2_{\rm min}$ as the area of the BAO surveys is increased, leading to mean $\chi^2_{\rm min}$ values that are greater than 5 (i.e. a reduced $\chi^2$ greater than one) for all survey configurations with 100\% of the DEFT stage III survey area. This is true with and without SNe systematics. 
In Figure \ref{fig:sne_num}, we show
similar results as a fraction of future SNe surveys. We see linear rise in $\langle\chi^2_{\rm min}\rangle$ with the SNe fraction from 1\% to 100\% of stage III experiments. Here, the rise is less dramatic than in the BAO case, and this suggests that it is more likely for discovery to come from the BAO experiment. This result can also be seen in Table \ref{tbl:errors}, where we also show the comparison with the other figures of merits discussed in sections \ref{sec:stdmod} and \ref{sec:ext_stdmod}. The middle column shows the errors on the standard model parameters, in this case the density of $\Lambda$, and on the right we show the FoM proposed by the DETF, which is proportional to the area of the error ellipse in the $w_0$-$w_a$ plane \cite{2006astro.ph..9591A}. Reassuringly, all three measures show similar trends, which would suggest that the simple optimisations done here are reasonably robust and the overall information content is increased between experiments with lower FoM and ones with higher ones. This is different from the tradeoff studied in appendix \ref{Illustrative Example}, where the overall error bars are fixed and the sensitivity in different regions ($x$ values) leads to changes in the FoMs.

\begin{figure}
\centering
\resizebox{\columnwidth}{!}{\includegraphics [angle = 0] {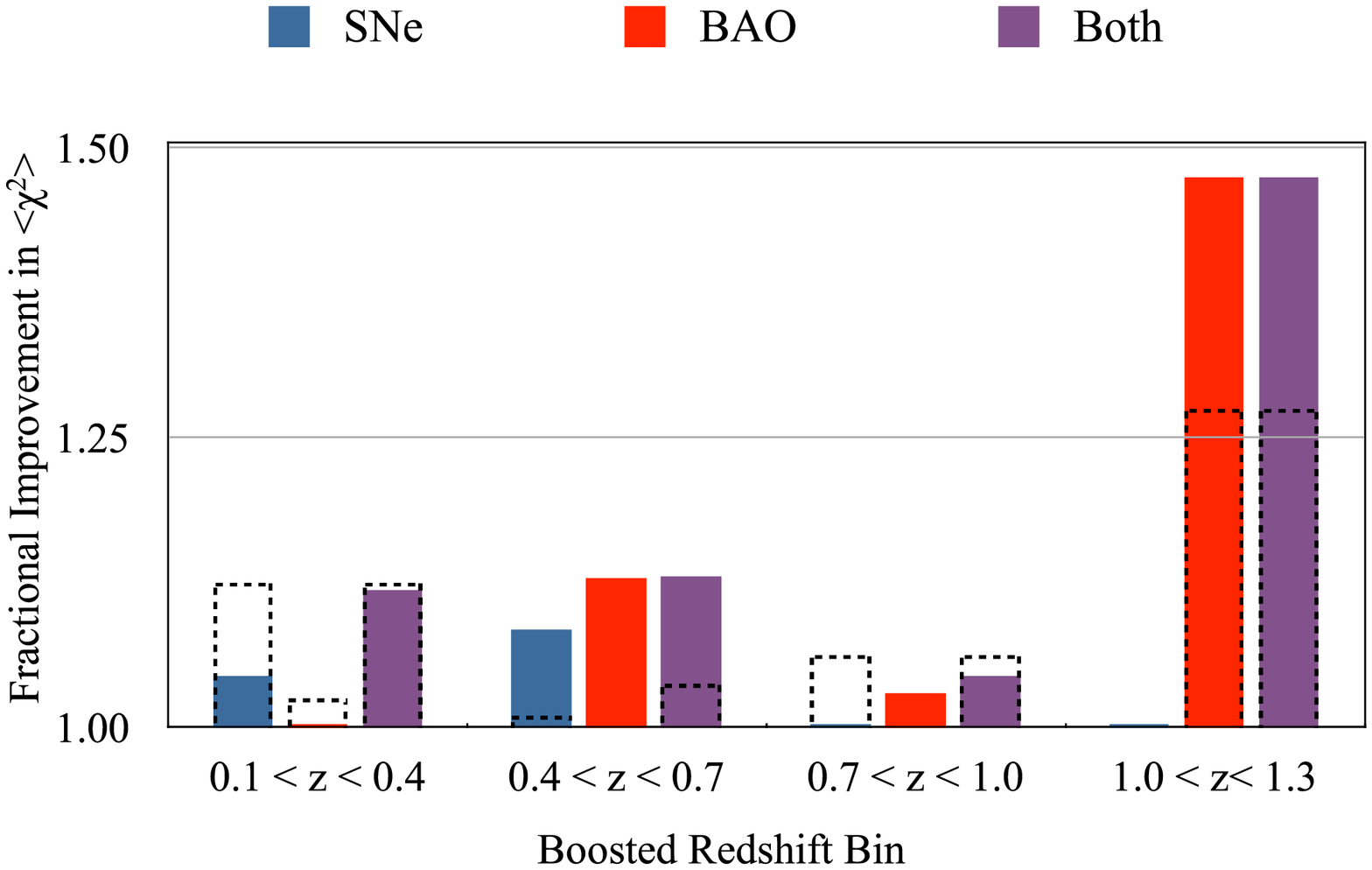}}
\caption{The impact of boosting the performance in one of the
  redshift bins. This is done by reducing the statistical error in the
  relevant bin by a factor of 2. The y-axis shows the ratio of the
  expectation value of the minimum $\chi^2$ of the boosted stage III
  survey relative to the standard stage III survey. The different
  colours corresponding to which probe have been enhanced; the solid
  colours are when SNe systematics are included, and the dashed lines
  show the results when SNe systematics are eliminated.} 
\label{fig:boost} 
\end{figure}

Finally, we investigate a simple optimisation where we explore the
model breaking redshift sensitivity of the Stage III surveys. We do
this by boosting the performance of the surveys at a particular
redshift by dividing the statistical errors at that redshift by a
factor of $2$. This is not a physically motivated optimisation. Instead, it can be thought of as simply probing where an improvement would be the most effective. The results are shown in Figure
\ref{fig:boost}. The coloured bars show the fractional increase in
$\langle\chi^2_m\rangle$ for the calculations where SNe systematics
have been included.  We see here that improving the SNe survey in the two lowest redshift bins causes a notable increase in the $\langle \chi^2_{\rm min}\rangle$, while improving the SNe performance in higher redshift bins has little effect, except in the no systematics case. This suggests that to go beyond stage III SNe experiments we should focus on improving errors at low redshifts first, unless we can demonstrate that the systematics levels can be brought below those presented by \cite{2004MNRAS.347..909K} and \cite{2006PhRvD..74d3513I}.  For the BAO experiments, we find a different result. Improving the errors in our lowest redshift bin has no effect on $\langle \chi^2_{\rm min}\rangle$. However, we see that if the errors in our final redshift bin ($0.7 < z < 1.0$) are improved, then we see the largest rise in $\langle \chi^2_{\rm min}\rangle$. This suggests that a BAO experiment beyond stage III should aim to make measurements at high redshifts.

\begin{table}
  \centering 
\begin{tabular}{l|c|c|c}
\hline
   & $\langle \chi^2_m \rangle$  & $\sigma_c(\Omega_\Lambda)/ \sigma_{III}(\Omega_\Lambda)$ & ${\rm FoM_{III} /FoM_c}$ \\
   \hline
  BAO III & 5.5 & 10 & 2.2\\
   SNe III & 3.5 (4.0)  &6 (14)  & 1.1 (2.6)\\
   BAO \& SNe III &  7.0 (8.0)  & 10 (19) &2.2 (4.4)\\
\hline
\end{tabular}
  \caption{Comparison between the model breaking approach
    ($\langle\chi^2\rangle$), working within the standard model (here
    we show errors on $\Omega_\Lambda$ in a model with only
    cosmological constant) and DETF FoM (which involved parameterising
    the equation of state in terms of $w_0$ and $w_a$). The numbers in parentheses are when no systematics are included for SNe, while the other numbers have this systematic included.} 
  \label{tbl:errors}
\end{table}

\section{Conclusions}
\label{Conclusion}

We have presented a framework in which experimental optimisation can be 
placed. Given a standard model, one can either (i) measure the standard model parameters
to high precision; (ii) attempt to extend the standard model; or (iii) attempt to 
find deviations from the standard model. 

When designing an experiment to measure or extend the standard model, the Fisher matrix formalism can be used. We have introduced a
framework that can be used to design an experiment to have the best
chance of finding discrepancies with the standard model. This framework only depends on three sets of information (current data, 
future expected error bars and the standard model). No external
assumptions are needed for the calculations, though we have also
shown how priors from the theory can, if needed, be added.  

By using a simple illustrative example, we find that the optimal future
experiment configuration can depend very strongly on the choice of
optimisation metric.  In our simple model, $C=m(x-8)+10$, the data
position $x=8$ is a pivot point since $C(x=8)\equiv 10$. When
designing an experiment to measure the standard model, it is optimal to
have small errors away from the pivot point. However, when designing an
experiment to break the model, it is optimal to have a small error at
the pivot point since any measurement of $C(x=8)\not=10$ would provide
evidence that the standard model was incorrect. When extending the
model, the optimisation naturally depends on the exact parameterisation
of the extension.  

In cosmology we have a standard model, $\Lambda$CDM. A large number of
experiments have been designed to measure an ad hoc extension of this
model, parameterisations of the dark energy equation of state, to high
accuracy. Our recommendation here is that future cosmology missions
should be optimised by using the three approaches we have outlined
above: (i) measure the standard $\Lambda$CDM parameters; (ii) measure
extended parameters, specifically the equation of state parameters,
the DETF FoM and the modified gravity parameter $\gamma$; and (iii) 
calculate the expectation value that the experiment will find a
deviation from $\Lambda$CDM. We calculate quantities in these three
regimes for SNe, BAO (transverse modes) and the CMB peak positions by
focusing on `current' and the DETF stage III surveys. Should the three
quality quantifiers agree, then we can be reassured that the
optimisation is somewhat robust. For instance, there has been some
concern that the DETF FoM is biased in favour of redshifts. However, in
the calculations shown in this paper, we do not find
evidence for this, with the results for the DETF FoM being consistent
with the other figures that we have shown. 
In the event that the three approaches lead to conflicting configurations, the the fact that these measures look for distinctly different thing means that we should be able to make a choice based on  a judgement of the priorities of a given experiment.

\section*{Acknowledgments}
We would like to thank Alexandre Refregier for the useful and
insightful discussions that initiated this work. We also thank Andy Taylor, Fergus Simpson and Anais Rassat
for useful comments. AA is supported by
the Zwicky Fellowship at ETH Zurich. TDK is supported by STFC rolling
grant number RA0888.  

%\newpage
\appendix

\section{Illustrative Example}
\label{Illustrative Example}

To illustrate the distinction between the three optimisation approaches 
highlighted in this article, we will present a simple worked example. 
We begin with a standard model where the signal $C$ at $x$ depends only on 
the parameter $m$ (i.e. $\Theta = \{m\}$). Our standard model is that 
\begin{equation}
\label{eq:SM}
C = m (x - 8) + 10.
\end{equation}

\begin{figure}
\centering
\resizebox{\columnwidth}{!}{\includegraphics{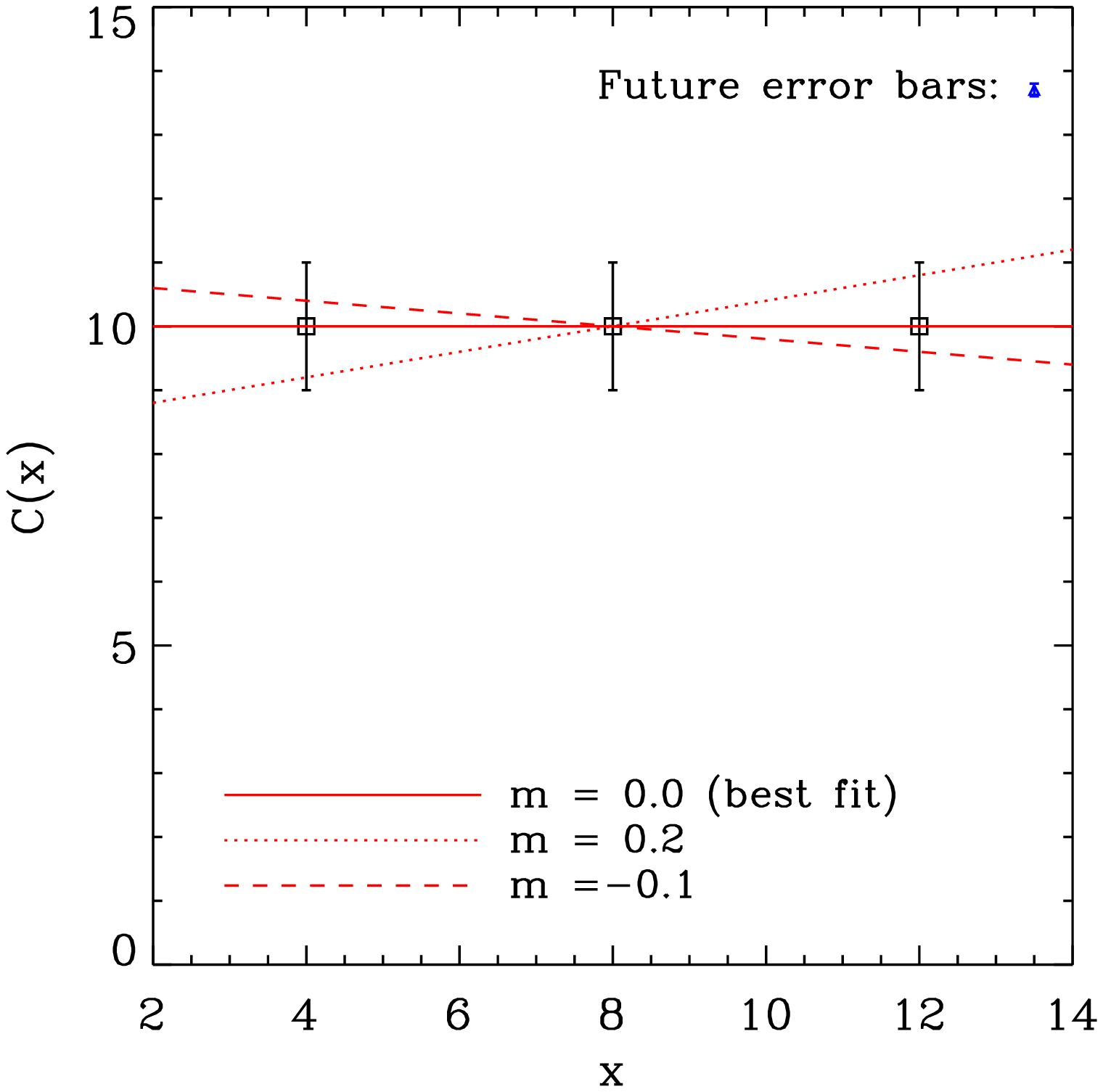}}
\caption{The system being used to illustrate the
  available optimisation options. For this example, the black points
  are today's data and the red lines are examples of our standard model
  that are consistent with today's data. In the top right hand corner,
  an example of the typical size of the error bars in the future
  experiment is shown.} 
\label{fig:ill1} 
\end{figure}

For this simple example, we also assume that measurements can only be made at 
$x= \{4,8, 12\}$, where today's measurements have yielded $X = \{10,10,10\}$ 
with Gaussian errors of variance $\sigma^2_Y = \{1,1,1\}$. 
This is shown in Fig. \ref{fig:ill1}. We will assume that future experiments 
can be built to measure the signal at the same $x$ positions as today but 
that the errors on the measurements will be significantly smaller than those of 
today. Specifically, we will assume that the quadratic sum of the future 
errors, over all data points, is $\sum_x \sigma^2_Y(x)=2.01$ (this creates a symmetry between the top left and right corners of Figures \ref{fig:MSM}, \ref{fig:ESM1},  \ref{fig:ESM2} and \ref{fig:BSM}). The global performance of the future experiment is, therefore, a little better than the current one, and the optimisation process is to decide how to optimally distribute the errors among the three data points.

\subsection{Measuring the Standard Model}

To measure performance of a future experiment, we use the Fisher matrix to 
estimate the errors on the parameter $m$ for specific configurations of the 
errors. This allows us to find the optimal configuration of the errors 
for the purpose of measuring $m$. 

\begin{figure}
\centering
\resizebox{\columnwidth}{!}{\includegraphics{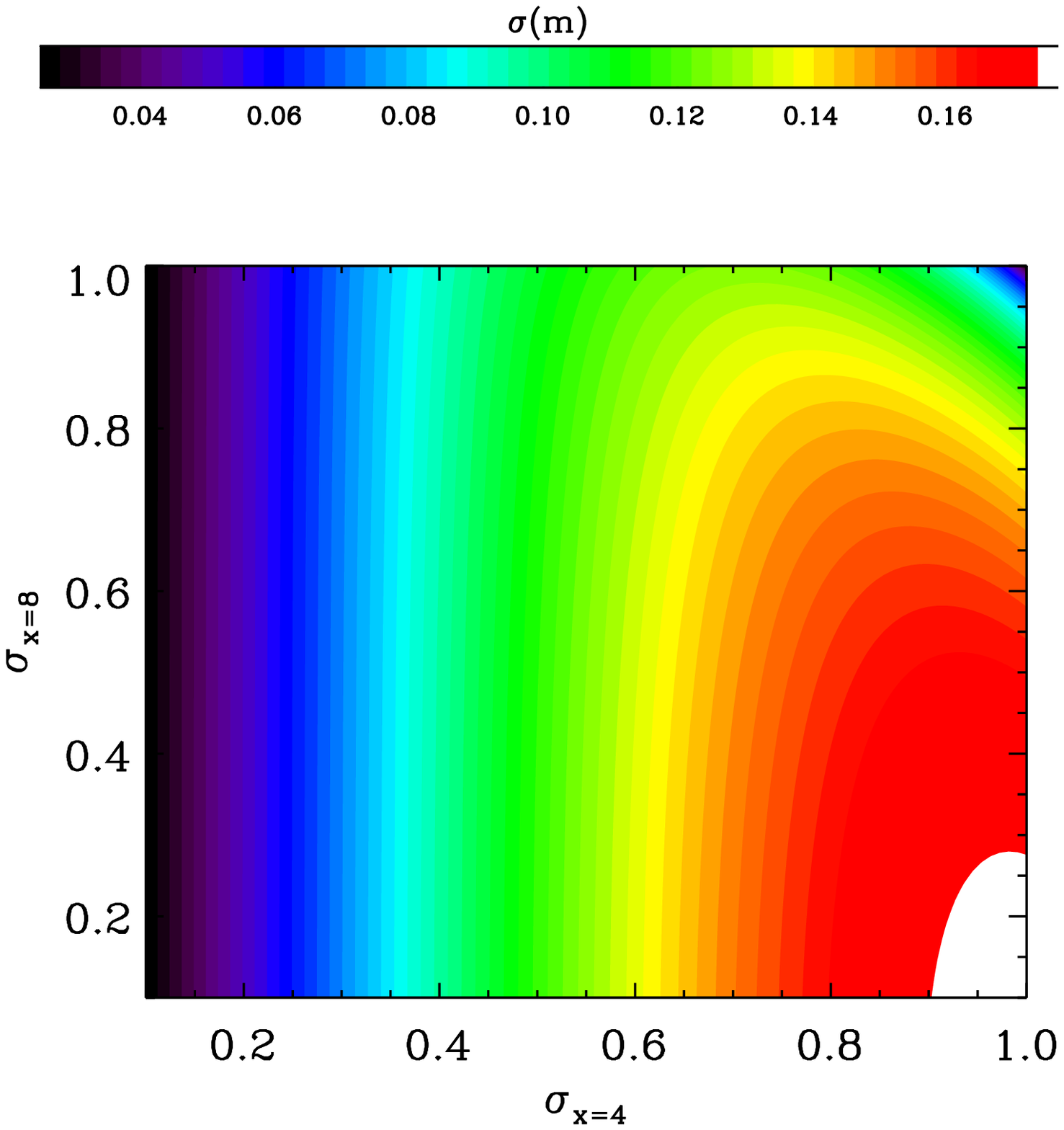}}
\caption{The results of an optimisation analysis designed to measure $m$ 
(the only parameter of our standard model) to the highest precision possible. 
The quadratic sum of the errors of the three data points, 
$(\sigma^2_{x=4} + \sigma^2_{x=8} +\sigma^2_{x=12})$, has been set to $2.01$. 
We see that the minimal errors are achieved for small $\sigma^2_{x=4}$ 
(and by symmetry $\sigma^2_{x=12}$). The fact that the lines are close 
to vertical shows that this optimisation is totally insensitive to the 
measurement precision at x=8. This can be understood since $x=8$ is 
a pivot in our standard model and therefore offers no information within 
our standard model since it can only have a value of $10$. For a fixed error 
at $x=8$, we see a clear preference to mimise the errors at either $x=4$ 
\emph{or} $x=12$, which means that it is better to have one small error bar than 
mimising both.}
\label{fig:MSM} 
\end{figure}

Fig. \ref{fig:MSM} shows how the future errors at the 
$x=4$ and $x=8$ points are optimised such that the error on 
$m$ is minimised. It is clear that the optimal configuration is insensitive 
to the error at $x=8$. This is understandable since within the standard 
model there is not sensitivity to $m$ at $x=8$, so there is no gain 
in placing any measurement at this point. The optimal strategy to measure 
the standard model $m$ is then to place small future error bars at either 
$x=4$ or $x=12$. It is also interesting to note that since the value of
the standard model at $x=8$ is fixed, it is better to have one small
error on either $x=4$ or $x=12$ (with the other being large) than to
distribute the errors between these two points. 

\subsection{Extending the Standard Model}

To extend the model, we first have to decide on a way of extending the standard 
model. We must also decide whether to optimise or minimise the 
errors on the extended parameters - after marginalizing over $m$ - or to 
minimize both the standard and extended parameters simultaneously. 

For illustration, we assume that there are two equally valid ways of extending the standard model 
used here. The first is the addition of a quadratic term,
\begin{equation}
C = m (x - 8) +10 +  p_1 (x - 8)^2, 
\end{equation}
and the second is the addition of a constant,
\begin{equation}
C = m (x - 8) +10 +  p_2.
\end{equation}
Again the Fisher matrix formalism is used 
to predict the future errors on the model parameters, ($m$, $p_1$) or 
($m$, $p_2$), given a configuration for the future data error bars. 

\begin{figure} 
\begin{center}
\resizebox{\columnwidth}{!}{\includegraphics{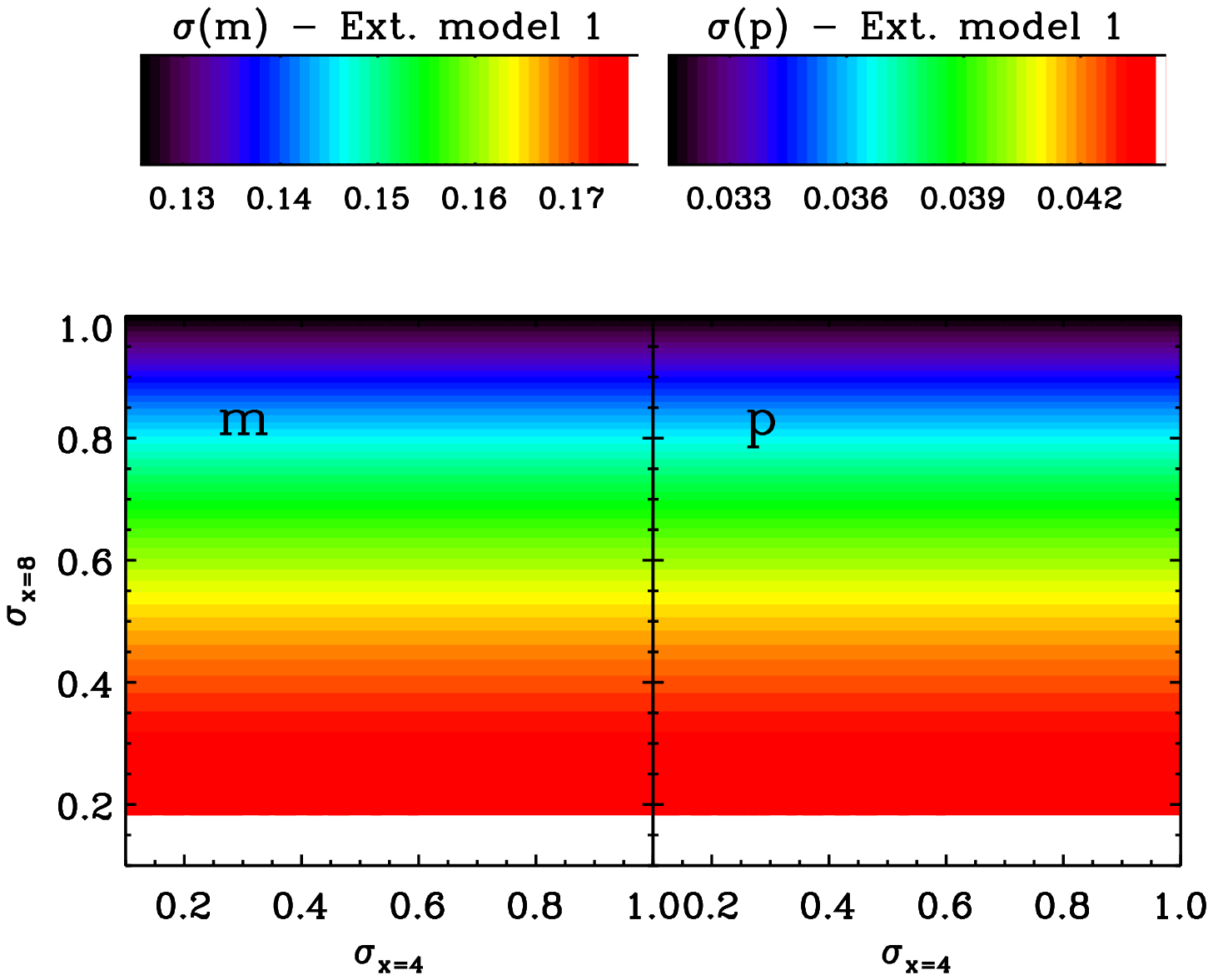}}
\caption{Optimisation for the two parameters of extended model $1$, 
$C = m (x - 8) +10 +  p_1(x - 8)^2$. The plots show the expected marginalised 
errors on $m$ and $p_1$ as a function of possible measurement errors at 
$x=4$ and $x=8$. As in Fig. \ref{fig:MSM}, the errors at $x=12$ are set by 
fixing the quadratic sum of the errors to $2.01$. For this extended model, 
we see that we are pushed to a configuration with maximum errors at $x=8$ for 
both parameters $m$ and $p_1$. As with the standard model, this extended 
model has a pivot point at $x=8$ and so the measurment here does not bring 
useful information. Unlike the example shown in Fig. \ref{fig:MSM}, here the 
errors at both $x=4$ \emph{and} $12$ are important since they are both needed to 
distinguish between $m$ and $p_1$ (with only one data point the two 
parameters are degenerate). This is why maximising the error at 
$x=8$, and hence minimising the quadratic sum at $x=4$ and $12$, is preferred.}
\label{fig:ESM1}
\end{center} 
\end{figure}

\begin{figure} 
\centering
\resizebox{\columnwidth}{!}{\includegraphics{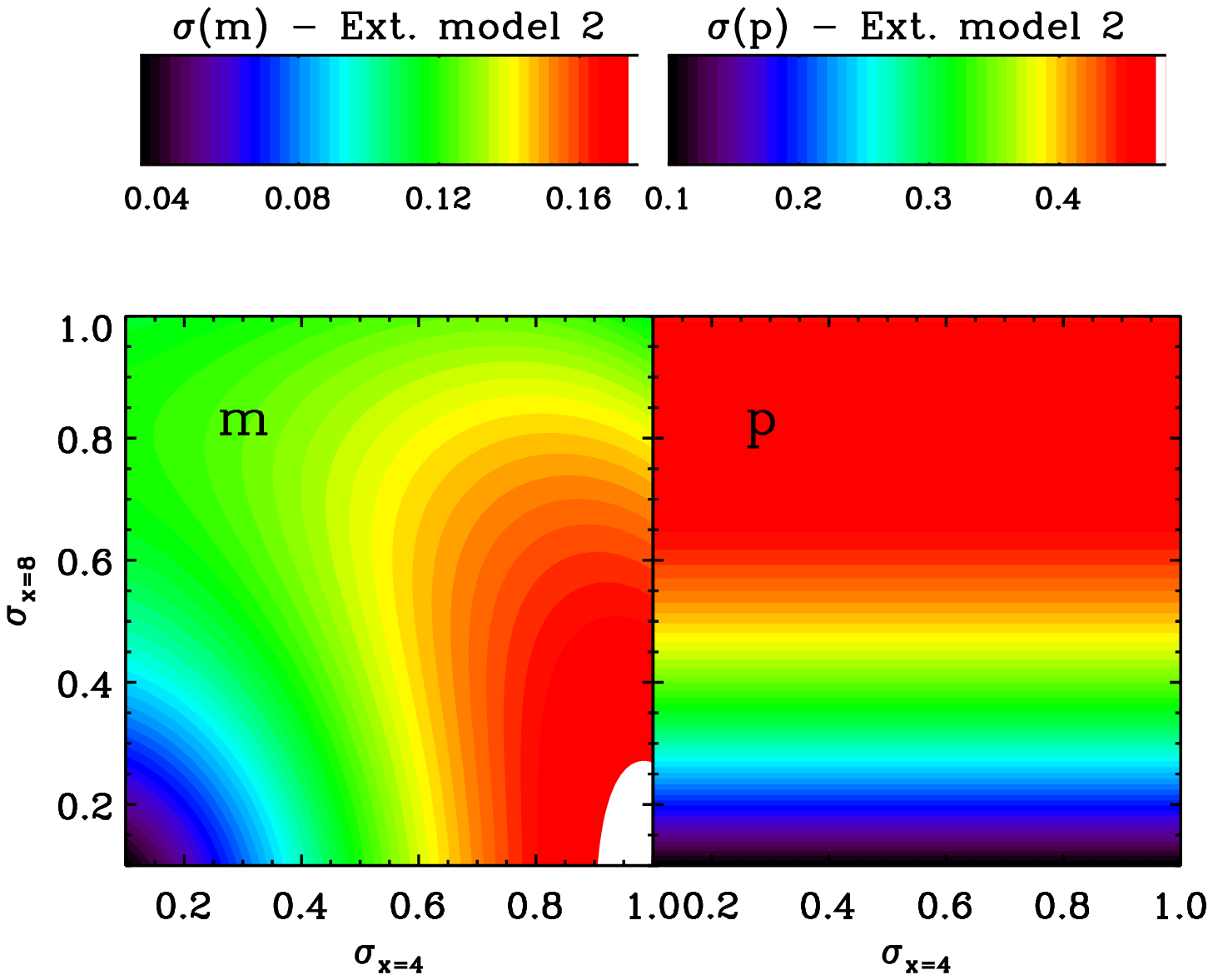}}
\caption{Similar to Fig. \ref{fig:ESM1}, this shows the optimisation for the 
two parameters of extended model $2$, $C = m (x - 8) +10 +  p_2$. 
For the extended parameter $p_2$, we are pushed towards a configuration with 
minimal errors at $x=8$. Because this point is not a pivot point of 
the model, it can be used to directly measure $p_2$. For $m$ we see 
that maximum precision is reached by minimising the errors at $x=4$ and $8$ 
(or by symetry at $x=8$ and $12$). This is because the data point at $x=8$ 
gives the best measure of $p_2$, which is degenerate with $m$. Once $p_2$ 
is measured, only one extra data point is needed in this model. Hence, either 
$x=4$ or $12$ should be minimised.}
\label{fig:ESM2} 
\end{figure}

The results are shown in Figs. \ref{fig:ESM1} and \ref{fig:ESM2}. 
We show the errors on $m$ (marginalised over $p_i$) and on $p_i$ 
(marginalised over $m$). We could have constructed a figure of merit that 
combines the errors of $m$ and $p_i$, but this is somewhat superfluous in this illustrative example. 

In Fig. \ref{fig:ESM1}, we show how the errors on $m$ and $p_1$ from model $1$ 
are optimised. In this case, $x=8$ is a pivot point of the extended model 
so the optimal strategy is to maximise future errors at $x=8$ since the 
parameters are not sensitive to data at this point. The quadratic sum of 
the errors at $x=4$ \emph{and} $12$ are then minimised. We note that in this example both of these data points are needed to distinguish between the parabolic and the linear term.

In Fig. \ref{fig:ESM2}, we show how the errors on $m$ and $p_2$ from model $2$ 
are optimised. In this extended model, $x=8$ is no longer a pivot point of the 
model. In fact, a small future error bar at $x=8$ could measure $p_2$ 
very accurately (for a given $m$) because the errors are not degenerate with $m$ at this point. Hence, the optimisation places a 
small error bar at $x=8$. Next, to accurately measure m, the
  optimisation tries to minimise the errors on {\it one} of the two
  remaining errors in a similar way to what happen in
  Fig. \ref{fig:MSM}.

\subsection{Breaking the Standard Model}

For the Fisher matrix calculations we have made the implicit assumption that future measurement errors are Gaussian \citep{1997ApJ...480...22T}. For the model breaking approach, we make the same assumptions, namely that the probability of T given today's 
data is given by
\begin{equation}
P(T|X) = \exp\Big({-\frac{(T-X)^2}{2\sigma_X^2}}\Big),
\end{equation}
where today's data vector is once again, $X = \{10,10,10\}$, and the probability of the future data given T is
\begin{equation}
P(Y|T) = \exp\Big({-\frac{(Y-T)^2}{2\sigma_Y^2}}\Big).
\end{equation}
The future $\chi^2$ is given simply by
\begin{equation}
\chi^2 = \sum_i \frac{1}{\sigma_{Y_i}^2} (C_i - Y_i)^2,
\end{equation}
which for the illustrative standard model used here (equation \ref{eq:SM}) 
is a minimum for
\begin{equation}
m=\frac{\sum_i \sigma_i^{-2} (x_i -8)(10 - Y_i)}{\sum_i \sigma_i^{-2} (x_i -8)},
\end{equation}
where the sums are over $x=4$, $8$, $12$. 
These allow us, for the simple model being considered here, to solve 
equation \ref{eq:chi} analytically. 

\begin{figure} 
\centering
\resizebox{\columnwidth}{!}{\includegraphics{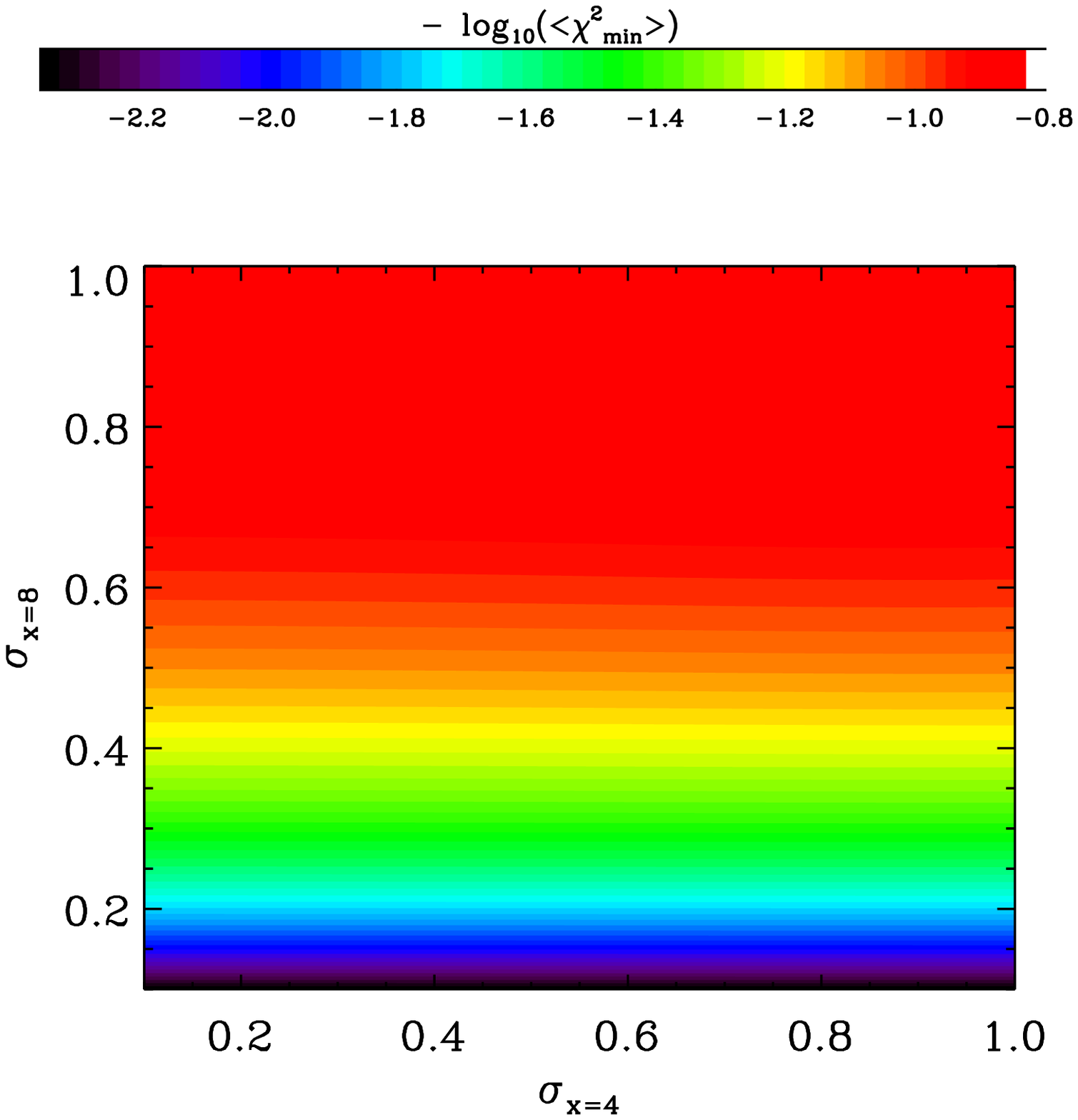}}
\caption{The expectation value of the future $\chi^2_{\rm min}$. 
This expectation value must be maximised to have the best chance of 
breaking our standard model. The colour scheme for this plot has been chosen 
such that the best configuration 
(${\rm max}(\langle \chi^2_{\rm min}\rangle)$) is purple (dark), which is 
consistent with Fig. \ref{fig:MSM} to \ref{fig:ESM2} where the optimal 
strategies are also purple (dark). We see that using this criterion that the 
optimal configuration is one that minimises the errors at $x=8$. This can be 
understood since {\bf any} deviation from $y\equiv 10$ at this point cannot 
be explained within our standard model. Given today's data and no 
guidance from theory, a high precision measurement here is, therefore, most 
likely to break the standard model.}
\label{fig:BSM} 
\end{figure}

Fig. \ref{fig:BSM} shows the result of the model breaking optimisation for 
this illustrative example. To have the best chance of breaking this standard 
model, one should place a very small error bar at $x=8$. This is understood 
since $x=8$ has a very stringent prediction that $C(x=8)\equiv 10$, \emph{any} 
deviation from this prediction would be proof that the standard model was 
incorrect.

\bibliographystyle{mn2e}
\bibliography{/Users/aamara/Work/Mypapers/mybib}

\end{document}